\documentclass[12pt]{article}
\usepackage[utf8]{inputenc}
\usepackage{graphicx} % Allows you to insert figures
\graphicspath{{tikz/}}
\usepackage{amsmath} % Allows you to do equations
\usepackage{fancyhdr} % Formats the header
\usepackage{geometry} % Formats the paper size, orientation, and margins
\usepackage{subcaption}
\usepackage{listings}
\usepackage{pdfpages}
\usepackage[english]{babel}
\usepackage{lscape}
\usepackage{tikz}
\usepackage[european,siunitx]{circuitikz}
\usetikzlibrary{arrows,calc,positioning}
\usetikzlibrary{angles, quotes}
\usepackage{pgfplots}
\pgfplotsset{compat=1.12}
\usepgfplotslibrary{external,polar}

\usepackage{cite}
\def\BibTeX{{\rm B\kern-.05em{\sc i\kern-.025em b}\kern-.08em
		T\kern-.1667em\lower.7ex\hbox{E}\kern-.125emX}}

\begin{document}
\begin{titlepage}
   \begin{center}
        \vspace*{5cm}

        \Huge{Examination Minutes}

        \vspace{0.5cm}
        \LARGE{Measurement of Single-Antenna-Element Beamforming}
            
        \vspace{3 cm}
        \Large{Chair of Information and Coding Theory\\(Christian-Albrechts-Universität zu Kiel)\\ and\\Institute for Microwave and Wireless Systems (Leibniz Universität Hannover)}
       
        \vspace{0.25cm}
        \large{Nils L. Johannsen, Lukas Grundmann}
       
        \vspace{3 cm}
        \Large{}
        
        \vspace{0.25 cm}
        \Large{Measurements conducted the 17th of January, 2022}

       \vfill
    \end{center}
\end{titlepage}

\setcounter{page}{2}
\pagestyle{fancy}
\fancyhf{}
\rhead{\thepage}
\lhead{\textit{Minutes of the Metrological Verification of Single-Antenna-Element Beamforming}}

\section*{Preface}
This document shall provide all knowledge gained in conjunction and preparation with the conducted measurements in the antenna measurement chamber of the Institute of Microwave and Wireless Systems (IMW) of the Leibniz University of Hannover (LUH). The measurements have been prepared and conducted by Lukas Grundmann, IMW, and Nils L. Johannsen, Chair of Information and Coding Theory (ICT) of the Christian-Albrechts-University (CAU) of Kiel. This minute shall allow a simpler understanding and quicker reapplication of the required calibrations and system setup for the measurements of further antennas. 

\section{Preparation}
This section summarizes all steps conducted before the antenna measurements.
All steps towards a successful realization of the beamforming are named and partially discussed.

\subsection{Previous Considerations}
In order to perform a successful beamforming, phase coherent signals as well as known and steerable amplitudes have to be realized.
Differences in phase and amplitude have to be compensated in software, using a software defined radio (SDR).

\subsubsection{System Concept for the Measurements}
\begin{figure}[h]
    \centering
    \input{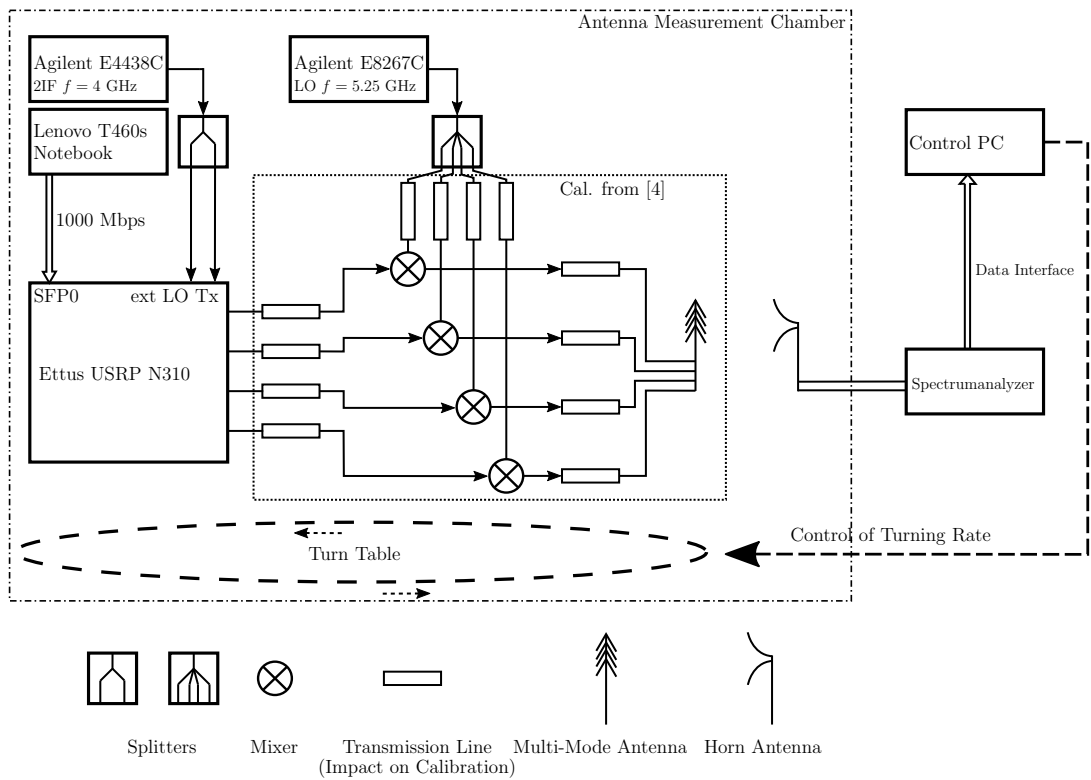}
    \caption{Block diagram of the measurement setup}
    \label{fig:MeasurementBlockDiagram}
\end{figure}

The antenna used for the omnidirectional optimization is the antenna specified in \cite{PeMa19a}.
It provides 6 antenna ports and is designed to work in the frequency band from 6-8.5\,GHz.
Due to hardware limitations by both project dependent use-case on a UAV as well as provided ports by the SDR, only four ports can be employed.
The port selection is done as follows:
The pattern has been optimized using a particle swarm optimization (PSO) to provide a close to omnidirectional gain pattern in the angular range between -45 and 45 degree, given the broadside direction with 0 degree.
The data of the far field of the antenna are drawn from field simulations of the model of the antenna under investigation.
Running the PSO for 50 times yields in an average power allocation at each port.
The two ports with the smallest power (ports number 1 and 3) are removed from the final set and the PSO is run again for the ports 2, 4, 5, and 6.
A SDR shall be used to apply the calculated precoding coefficients.
The SDR under consideration is an Ettus N310 SDR, which provides MIMO-capability of four ports in the frequency region between 10\,MHz and 6\,GHz \cite{EttusN310}.
Since the targeted frequency of the antenna is 7.25\,GHz, at the transmitter a hardware setup similar to the one presented in \cite{MAVMJ19} is chosen.
In Fig.~\ref{fig:MeasurementBlockDiagram}, a block diagram of the transmit path consisting of SDR, mixers, and antenna is depicted, as well as the receive path, employing a horn antenna, spectrum analyzer and computer for controlling both reception of signals and steering of the antenna angle.
Note that the horn antenna is capable of receiving both vertically and horizontally polarized signals.
Therefore, the horn antenna provides two output ports, one for each polarization.
Summarizingly, the transmit path consists of three major parts: The SDR, the mixing stage including transmission lines to the antenna, and the antenna including the feeding network itself. 

The hardware used for the setup is named in table~\ref{tab:RequiredHardware}.
\begin{landscape}
\begin{table}[h]
    \caption{Hardware required for measurement and calibration, as in \cite{Grundmann22}.}
    \label{tab:RequiredHardware}
    \centering
    \begin{tabular}{l|l|l}
     Function & Device & Explanation \\
      \hline
      Vector Network Analyzer & Rohde \& Schwarz ZVA 40 & Network analyzer for phase and amplitude measurements\\
      Signal Generator & Agilent E8267C & LO for calibration and measurement\\
      Signal Generator & Agilent E8267C & IF for calibration and SDR LO source\\
      Mixer 1 \& 2 & Marki M20243LP & Mixers connected to the SDR Ports 0 and 1\\
      Mixer 3 \& 4 & Marki M10220LA & Mixers connected to the SDR Ports 2 and 3\\
      Splitter & MiniCircuits ZFRSC-42-S+ & Splitter of the IF signal, for calibration and as SDR LO source\\
      Splitter & Marki PD40R526 & Splitter LO, calibration and measurement\\
      Transmission Line & Minibend R-7 & Connection from ports 1 and 2 of splitter IF\\
      Transmission Line & Minibend R-30 & Connection to port S of splitter IF\\
      Transmission Line & Minibend R-25 & Connection IF (Cal.) and SDR (Meas.) to mixers\\
      Transmission Line & Minibend R-4 & Connection LO to mixers\\
      Transmission Line & Gore 3GW40 0TD01D01048.0, 3.5\,mm-3.5\,mm & Connection to antenna\\
      Connector & Minibend R-10 & Connection to antenna transmission line\\
      
      $\ge$ 6 Terminations & $50~\Omega$-Terminations & Terminations for open antenna, splitter and VNA ports\\
    \end{tabular}
\end{table}
\end{landscape}

\subsection{Measurements to Prepare the Compensation of Hardware Impairments of Mixers and Transmission Lines}

As discussed above, the hardware between SDR and antenna consists of mixers and transmission lines.
Both have an impact on attenuation and phase.
To compensate this impact successively, all parts of the system are measured independently.
A minute about the measurements of mixer stage and transmission lines has been provided in \cite{Grundmann22}.
To accomplish a precise and complete description, parts of these notes will be included in this document.
When mixing the intermediate signals of the SDR to the desired transmission band, the mixers of the different channels may cause different attenuations and phase shifts.
Hence, the properties of each mixer and peripheral transmission lines have to be measured prior to the beamforming measurement.

\subsubsection{Measurement}
\begin{figure}[h]
    \centering
    \input{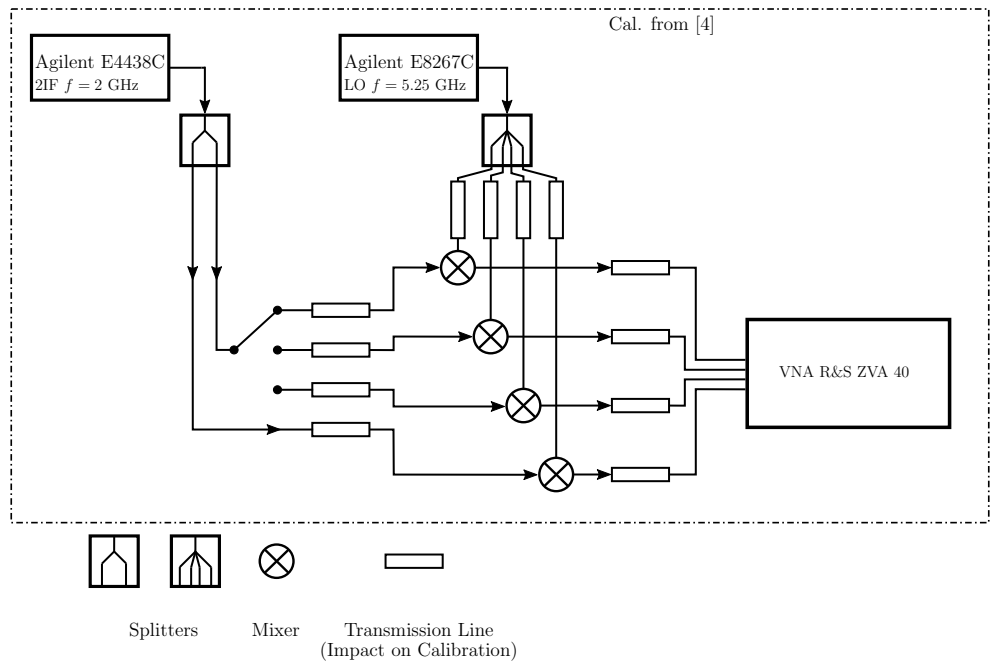}
    \caption{Block diagram presenting the measurement of network of mixers and transmission lines.}
    \label{fig:BlockDiagrammMixerTransmissionLinesCal}
\end{figure}

As aforementioned, this section is based on the description in \cite{Grundmann22}.
In Fig.~\ref{fig:BlockDiagrammMixerTransmissionLinesCal}, a block diagram representing the measurement of the network of mixers and transmission lines is depicted.
The network analyzer measures the ratio of wave quantities $b_p/b_1$, where the index $p$ denotes the number of port, e.g. 2, 3, and 4.
Since this measurement cannot be calibrated by the network analyzer itself, the different pathlengths need to be measured prior.
This measurement is done as follows: A signal generator provides a signal via a splitter to two ports of the VNA.
Since the splitter is known to have very small impact on phase and amplitude \cite{Grundmann22}, the resulting difference in attenuation and phase between two different paths of the VNA can be seen from the measurement result.
The phase has to be added during the measurement to yield valid results.

\subsubsection{Results for Calibration}
The results measured in \cite{Grundmann22} can be summarized as shown in table \ref{tab:HardwareComp}. The correction term is calculated by inverting the measured / determined values.
\begin{table}[h!]
    \centering
    \caption{Measured and Compensating Phases and Amplitudes of the Mixers and Transmission Lines, frequency $f=7.25~\textrm{GHz}$, results from \cite{Grundmann22}. According to the notes, the influence of the different path-lengths inside the VNA were compensated during the determination of the applicable correction factors.}
    \begin{tabular}{l|c|c|c|c}
        Port Number $p$ & 1 & 2 & 3 & 4\\
        \hline
        Ratios of wave quantities $b_p/b_1$ & 1 & $1.026\angle{162.30^{\circ}}$ & $1.186\angle{-158.24^{\circ}}$ & $1.248\angle{-138.24^{\circ}}$ \\
        Phase Correction & 0 & $-162.30^{\circ}$ & $158.24^{\circ}$ & $138.24^{\circ}$ \\
        Amplitude Correction & 1 & 0.9747 & 0.8432 & 0.8013 \\
    \end{tabular}
    \label{tab:HardwareComp}
\end{table}

\subsection{Compensation of the Feeding Network of the Antenna}

The measurement minutes in \cite{Grundmann22} already contain a table of the phases of the antenna feed network. 
The phases are taken from the field simulations conducted with the antenna model of the antenna described in \cite{PeMa19a}.
The phases are provided in table~\ref{tab:PhasesFeednetwork}, and need to be applied with inverted prefix by the SDR to be compensated.
\begin{table}[h]
    \centering
    \begin{tabular}{c|c}
       Port Number  & Differential Phase in Degree with Respect to Port 1 \\
       \hline
        1 & 0\\
        2 & 266\\
        3 & 157\\
        4 & 245\\
        5 & 7\\
        6 & 17\\
    \end{tabular}
    \caption{Phases with respect to antenna port 1 of the feeding network of the antenna, as provided in \cite{Grundmann22}.}
    \label{tab:PhasesFeednetwork}
\end{table}

\subsection{Setup and Synchronization of the Software Defined Radio}

\subsubsection{Setup}
The SDR is controlled using GNU Radio Companion (GRC), a free software development framework, which allows to control several SDR-devices.
For the control of the Ettus USRP N310, another software, USRP Hardware Driver (UHD), is required as a software interface between SDR and GRC.
In the context of this work, GRC version 3.9.4.0 and UHD version 3.15.0 are used.
The computer controlling the SDR is a Lenovo Thinkpad T460s, which is connected to the SDR via Ethernet.
This bottleneck limits the achievable data rates.
Hence, only low sampling rates are used during the measurements.
The advantage of using this connection in conjunction with the notebook is the ability to reduce the size of the reflecting surfaces in the antenna measurement chamber during the measurements in comparison to a larger workstation providing faster connectivity. 
Although the SDR provides MIMO capabilities employing its four Tx or Rx ports, out of the box it is not capable of generating phase-coherent signals and requires an external local oscillator (LO) for both calibration and in service operation \cite{EttusN310}.
The LO signal of frequency $f_{\textrm{LO,SDR}} = 4~\textrm{GHz}$ for the LO-inputs of the SDR is provided by a signal generator (Agilent E4438C).
The LO signal is split by a Minicircuits ZFRSC-4-842-S+ splitter.
The resulting signals at the outputs of the SDR provide half the frequency of the LO, hence $f_{\textrm{IF}} = 2~\textrm{GHz}$. 

\subsubsection{Calibration}
\begin{figure}[h]
    \centering
    \input{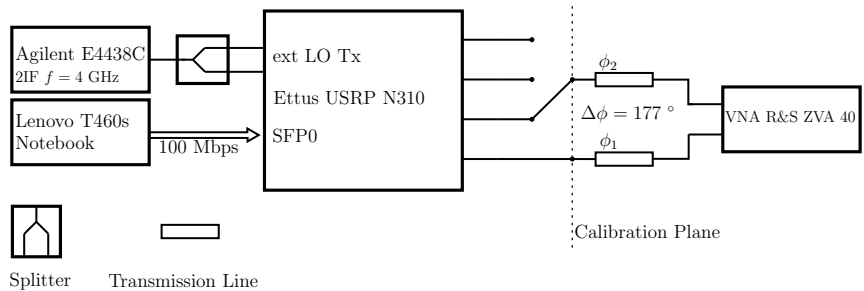}
    \caption{Blockdiagram of the calibration of the SDR. The phases of the transmission lines and their difference are shown for illustrational purposes. The phases are dependent on frequency. Of course, attenuation applies as well, but has rather small difference and therefore small impact on the beamforming. IMPORTANT!: Disconnected ports have to be terminated using $50~\Omega$ termination!}
    \label{fig:BlockDiagramSDRCal}
\end{figure}

Any phase differences of the two LO-signals are not of interest, since the resulting output signal of the SDR needs to be calibrated, which solves any unresolved phase differences in the LO-signals at the same time. 
The block diagram of the calibration is provided in Figure~\ref{fig:BlockDiagramSDRCal}.
Note that all disconnected transmit ports of the SDR are terminated using a $50~\Omega$-termination.
As for the measurement of the differential phases for mixers and transmission lines, a network analyzer is used for the calibration of the SDR. 
In a first step, the differential phase and attenuation of the two transmission lines connecting the vector network analyzer (VNA) R\&S ZVA 40 to the SDR are measured.
This is done by providing a sinusoidal signal from the Agilent E4438C signal generator via a splitter to two ports of the VNA. 
The phase difference of the transmission lines can be taken from the fraction of the wave quantities $b_2/b_1$.
In the measurement setup, a phase difference of $177^{\circ}$ has been measured.
The difference of attenuation of the wires can be disregarded, since it is fairly small.
For the calibration of the SDR, ports 0 and 1 of the SDR are connected to port 1 and 2 of the VNA.
Recall that all open transmit ports of the SDR have to be terminated using $50~\Omega$-terminations.
A picture of the calibration in the measurement chamber is given in Fig.~\ref{fig:CalibrationChamber}.
\begin{figure}[t]
    \centering
    \includegraphics[width=0.6\textwidth]{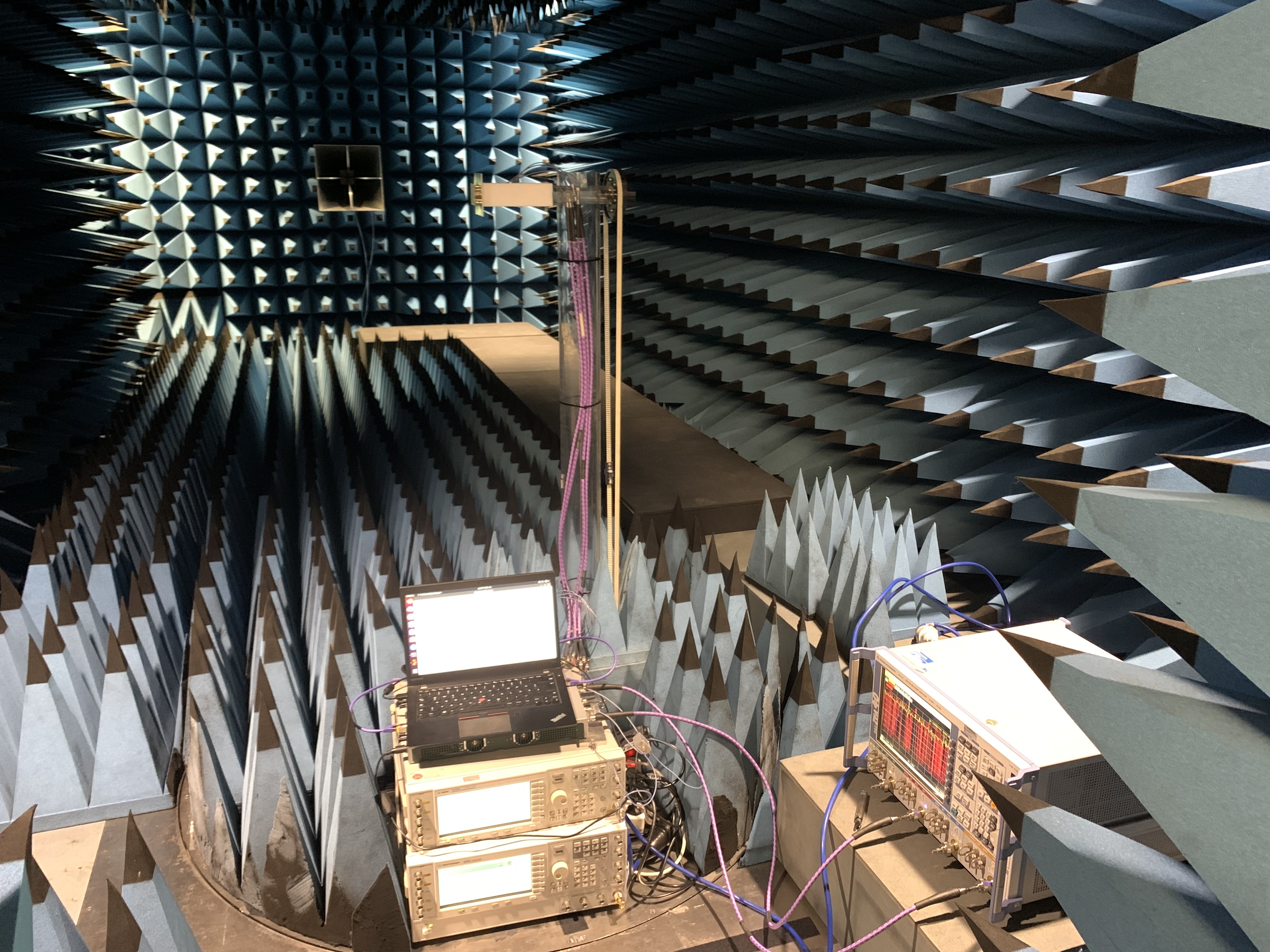}
    \caption{Calibration of the SDR in the measurement chamber. The notebook is located on the signal generators and SDR, the VNA at the right. In the top the multi-mode antenna can be seen, with the horn antenna in the background.}
    \label{fig:CalibrationChamber}
\end{figure}
\begin{figure}[h!]
  \begin{subfigure}[b]{\linewidth}
    \centering
    \includegraphics[width=\textwidth]{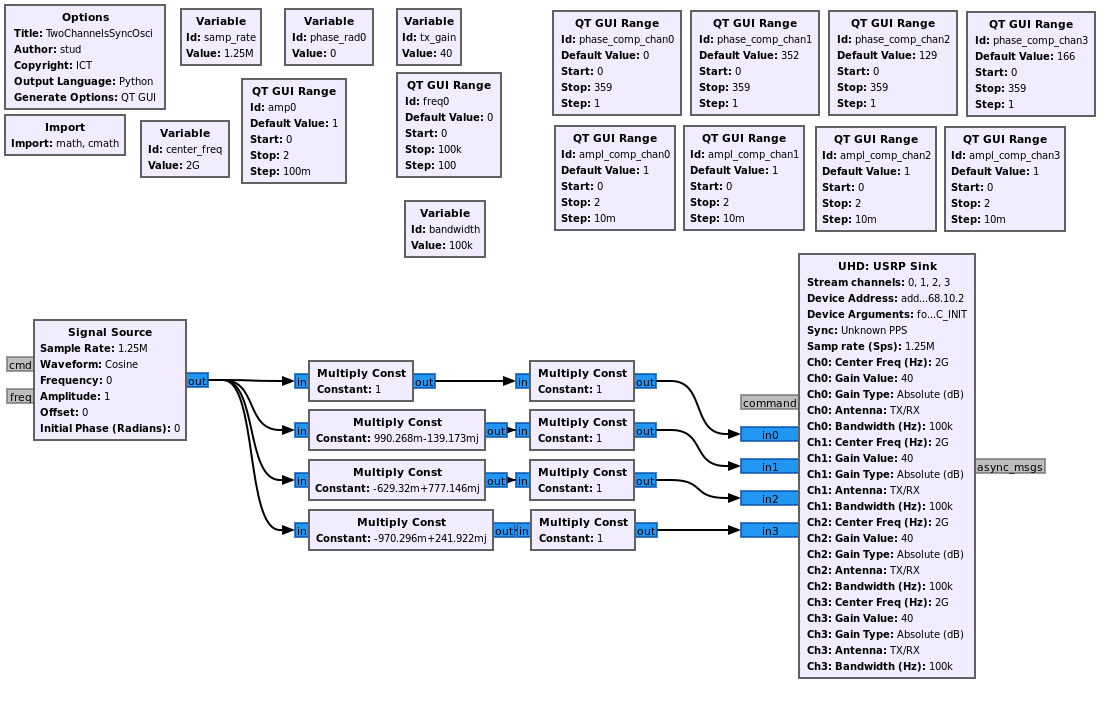}
    \caption{Flowchart of the Calibration Program.}
  \end{subfigure}
  \begin{subfigure}[b]{\linewidth}
    \caption{Listing of the Device Settings in GNURadio.}
    \label{lst:DeviceSettingsCal}
    \begin{lstlisting}
        # Device Address:
        "addr=192.168.10.2"
        # Device Arguments:
        "force_reinit=1, # Only for calibration
        rx_lo_source=external,tx_lo_source=external,
        init_cals=BASIC|TX_ATTENUATION_DELAY|RX_GAIN_DELAY|
        PATH_DELAY|LOOPBACK_RX_LO_DELAY|RX_LO_DELAY|
        RX_QEC_INIT|TX_QEC_INIT"
    \end{lstlisting}
  \end{subfigure}
  \caption{Software Settings for the Calibration of the SDR using GNURadio.}
\end{figure}
In the beginning of each run of the flowgraph, the SDR performs internal calibrations as defined by listing~\ref{lst:DeviceSettingsCal}, which is forced by the option "\textit{force\_reinit=1}".
When running these calibrations, the external LO should be set to 5~GHz, since only in this particular case the Quadrature Error Correction (QEC) calibrations can run successfully \cite{Poehlmann20}.
After a first run including the internal calibrations, the option "\textit{force\_reinit=1}" can be removed and the signal generator does not have to be set to 5~GHz anymore.
However, after the reduction of the LO frequency to the desired frequency of 4~GHz, the calibration of the ports can be done.
If the LO is disconnected from the SDR for any reason, phase ambiguities of $180^{\circ}$ apply \cite{Poehlmann20}. 
Using the SDR port 0 and VNA port 1 as a reference, all other ports of the SDR are connected to port 2 of the VNA iteratively.
In the phase of disconnection, the signal power of the current SDR port has to be reduced to zero to avoid damages due to reflected powers.
By modifying the amplitude and phase of the port under consideration, a power ratio of 0~dB and the aforementioned measured angle of $177^{\circ}$ can be set. 
The resulting calibration coefficients are noted in table \ref{tab:SDRCalibration}.
\begin{table}[h]
    \centering
    \caption{Coefficients for the calibration of the SDR.}
    \begin{tabular}{c|c|c}
       Port  & Phase in Degree & Amplitude \\
       \hline
        0 & 0 & 1 \\
        1 & 5 & 0.93 \\
        2 & 178 & 0.89 \\
        3 & 184 & 0.92 \\
    \end{tabular}
    
    \label{tab:SDRCalibration}
\end{table}
\newpage
 
\newpage
\section{Realization}
In this chapter, the realization of the measurement is described and the results are presented.
As a first step, the calibration needs to be applied.
Afterwards, the SDR and the notebook as well as any other devices are hidden underneath radiation absorbers on the turnable plate in the antenna measurement chamber.
Finally, the measurement is started from outside the locked chamber and paused any 90 degree.
The pause allows checking the pathway of the power cable which provides power to the devices on the turnable plate and is winded off during the rotation of the plate.
\begin{figure}[t]
    \centering
    \includegraphics[width=0.6\textwidth]{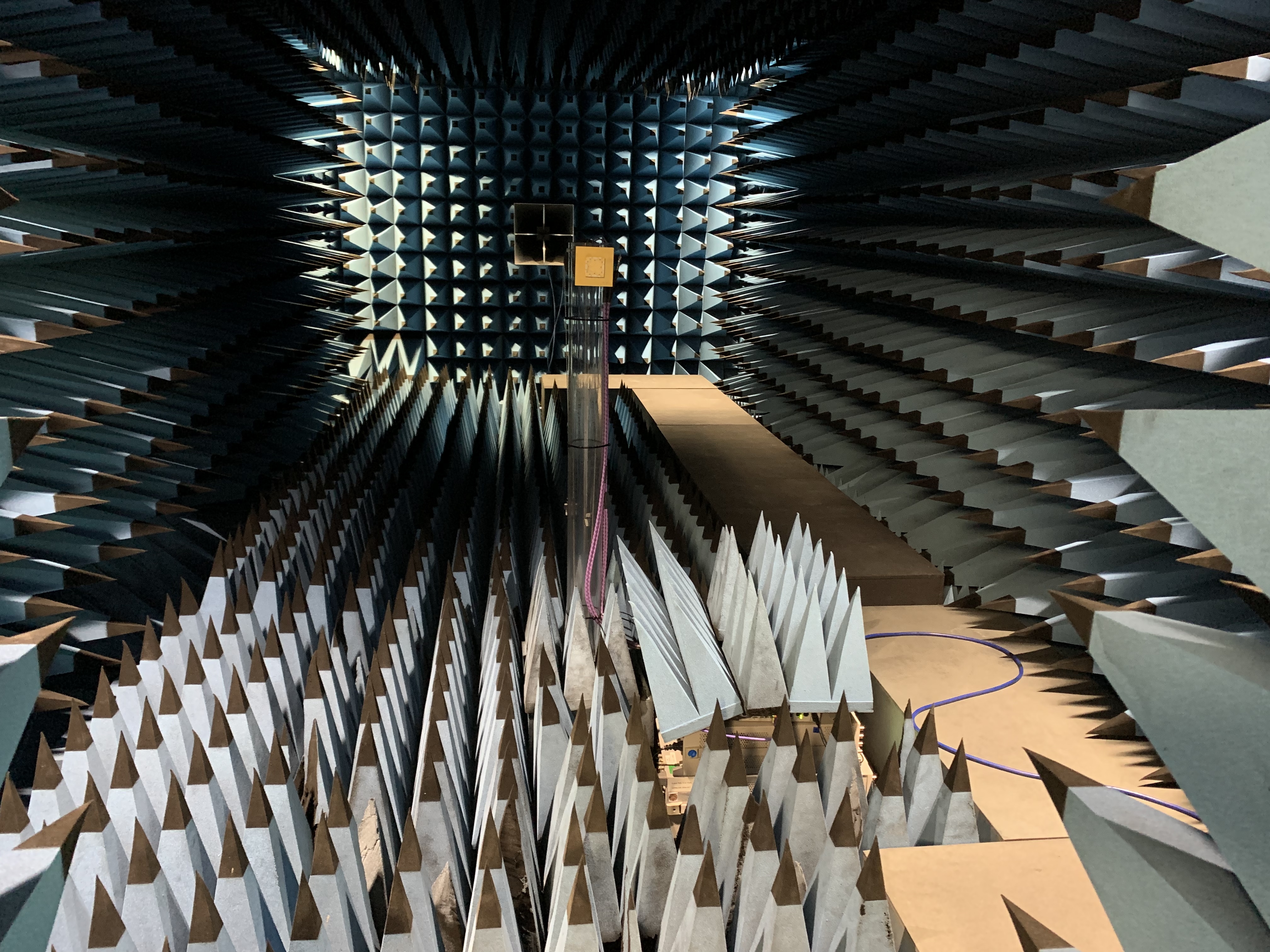}
    \caption{The setup is ready for measurement. As can be seen, the notebook is closed and the devices are covered by radiation absorbers.}
    \label{fig:CalibrationChamber}
\end{figure}

\subsection{Application of the Calibration}
\begin{landscape}
\begin{figure}[h]
    \centering
    \includegraphics[height=0.6\textwidth]{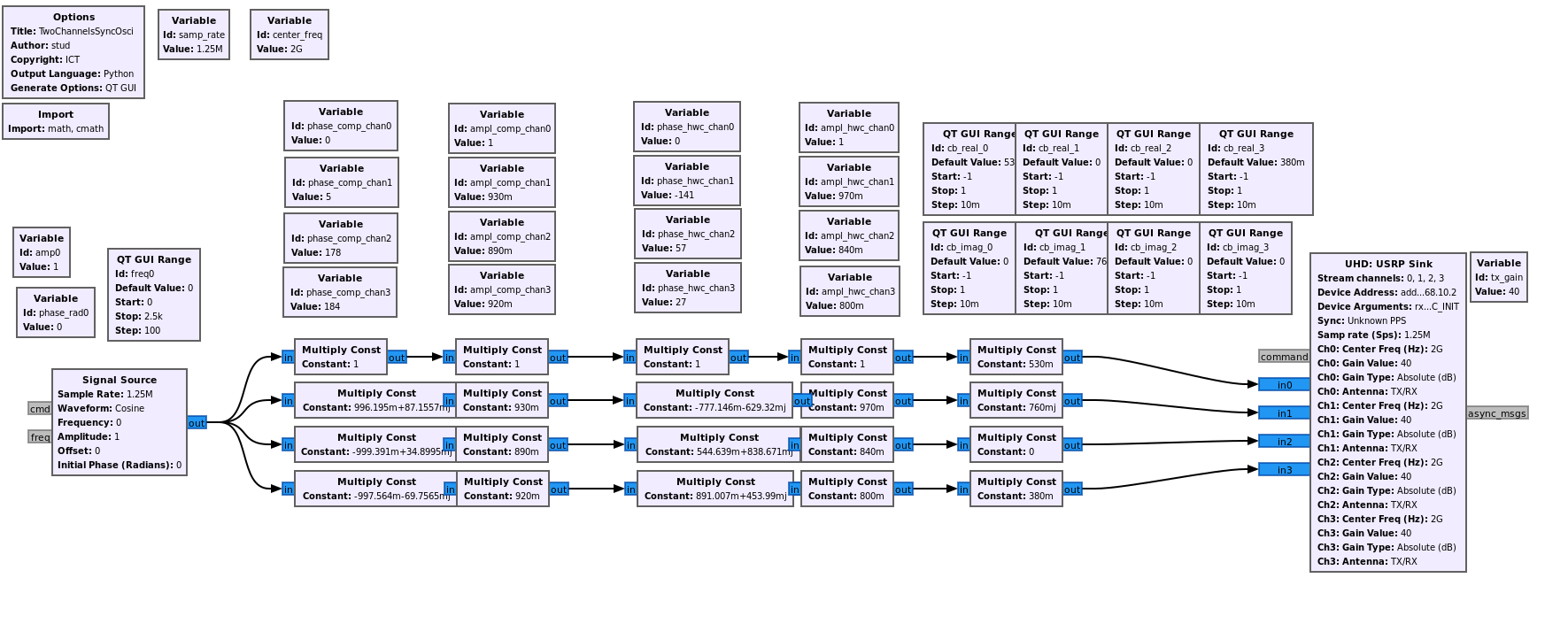}
    \caption{Flowchart of the Program for the Measurement in GNU Radio Companion.}
    \label{fig:FlowChartMeasurement}
\end{figure}
\end{landscape}

For the application of the calibration, a flowchart in GRC has been created. The flowchart is shown in Fig.~\ref{fig:FlowChartMeasurement}. 
The corresponding settings of the software were already given in listing~\ref{lst:DeviceSettingsCal},
and are the same as used for the calibration.
However, the parameter "\textit{force\_reinit=1}" is removed, since the already conducted calibrations shall be used by the SDR.
\subsection{Conducted Measurements}
\begin{figure}[t]
\centering
\begin{subfigure}[h]{0.65\textwidth}
  %{0.65\columnwidth}
\begin{tikzpicture}
    %\useasboundingbox (-1,-1.8) rectangle (15,8);
    %\draw [draw=black] (-1,-1.8) rectangle (15,8);
		\begin{polaraxis}[
		xmin = -90,
		xmax = 90,
		scale = 1,
		width = \columnwidth,
		xticklabel=$\pgfmathprintnumber{\tick}^\circ$,
		xtick={-90,-60,...,90},
	%	x coord trafo/.code=\pgfmathparse{#1+180},
	%	x coord inv trafo/.code= \pgfmathparse{#1-180},
	%	xtick={-90,-60,...,90},
		ytick={-10,-8,...,0},
		ymin=-10, ymax=1,
		y coord trafo/.code=\pgfmathparse{#1+10},
		rotate=-90,
		y coord inv trafo/.code=\pgfmathparse{#1-10},
		x dir=reverse,
		xticklabel style={anchor=-\tick-90},
		yticklabel style={anchor=east, xshift=2.5mm, yshift = -3mm},
		legend style={at={(0.6,0)}},
		legend columns = 4,
		ylabel = dB,
		y label style={at={(axis description cs: 0.54,0.18)},rotate=0},%,anchor=south},
	%	y axis line style={yshift=-4.75cm},
	%	ytick style={yshift=-4.75cm}
		]

		%,minGain,maxGain,meanGain,mpvarGain,mmvarGain
		
		\addplot[no markers, black] table [col sep =comma, x expr = {\thisrow{Angle}},  y expr ={\thisrow{G}}] {./tikz/GainPatternsOmni.csv};
	    \addplot[no markers, blue,thick] table [col sep =comma, x expr = {\thisrow{Angle}},  y expr ={\thisrow{s_vert}}] {./tikz/GainPatternsOmni.csv};
	    \addplot[no markers, green,thick] table [col sep =comma, x expr = {\thisrow{Angle}},  y expr ={\thisrow{s_horz}}] {./tikz/GainPatternsOmni.csv};
		\addplot[no markers, black, dashed] table [col sep=comma, x expr = {\thisrow{Angle}}, y expr ={\thisrow{GdB}}]{./tikz/gainPatternOmni.csv};
		
		\legend{\tiny{Gain}, \tiny{Vertically Polarized Pattern}, \tiny{Horizontally Polarized Pattern}, \tiny{Simulated Gain Pattern}}

		%\end{semilogxaxis}
	%	\legend{\footnotesize{Port 1}, \footnotesize{Port 2}, \footnotesize{Port 3}, \footnotesize{Port 4}}
		\end{polaraxis}
		 %\draw (current bounding box.north east) -- (current bounding box.north west) -- (current bounding box.south) -- (current bounding box.center) -- cycle;
      %   \tikzstyle{every path}=[draw] 
         \path
            (current bounding box.center)coordinate(C)
            (current bounding box.east)coordinate(E)
            (current bounding box.north)coordinate(N)
            ;
         \pgfresetboundingbox
        \path let
              \p1=(C),
              \p2=(E),
              \p3=(N),
              \n1={\x2-\x1}, % diff center-east
              \n2={\y3-\y1}, % diff center-north
              \n3={\x1-\n1}, % New West
              \n4={\y1+\n2}, % New North
              \n5={\x1+\n1}, % New East
              \n6={0-1/6*\y3} % New south
              in 
              [use as bounding box] (\n3,\n4) rectangle (\n5,\n6);
            
	%	current bounding box.south = current bounding box.south+(5,0);
	%	\draw (current bounding box.north east) -- (current bounding box.north west) -- (current bounding box.south east) -- (current bounding box.south west) -- cycle;
\end{tikzpicture}
  \caption{Omnidirectional}
\end{subfigure}
\begin{subfigure}[h]{0.65\textwidth}
  %{0.65\columnwidth}
\begin{tikzpicture}
   % \useasboundingbox (-1,-1) rectangle (6,3.5);
     %\useasboundingbox (-1,-1.8) rectangle (15,8);
    %\draw [draw=black] (-1,-1.8) rectangle (15,8);
		\begin{polaraxis}[
		xmin = -90,
		xmax = 90,
		scale = 1,
		width = \columnwidth,
		xticklabel=$\pgfmathprintnumber{\tick}^\circ$,
		xtick={-90,-60,...,90},
	%	x coord trafo/.code=\pgfmathparse{#1+180},
	%	x coord inv trafo/.code= \pgfmathparse{#1-180},
	%	xtick={-90,-60,...,90},
		ytick={-10,-8,...,0},
		ymin=-10, ymax=1,
		y coord trafo/.code=\pgfmathparse{#1+10},
		rotate=-90,
		y coord inv trafo/.code=\pgfmathparse{#1-10},
		x dir=reverse,
		xticklabel style={anchor=-\tick-90},
		yticklabel style={anchor=east, xshift=2.5mm, yshift = -3mm},
		legend style={at={(0.6,0)}},
		legend columns = 4,
		ylabel = dB,
		y label style={at={(axis description cs: 0.54,0.18)},rotate=0},%,anchor=south},
	%	y axis line style={yshift=-4.75cm},
	%	ytick style={yshift=-4.75cm}
		]

		%,minGain,maxGain,meanGain,mpvarGain,mmvarGain
		\addplot[no markers, black] table [col sep =comma, x expr = {\thisrow{Angle}},  y expr ={\thisrow{G}}] {./tikz/GainPatterns30deg.csv};
		\addplot[no markers, blue] table [col sep =comma, x expr = {\thisrow{Angle}},  y expr ={\thisrow{s_vert}}] {./tikz/GainPatterns30deg.csv};
	    \addplot[no markers, green] table [col sep =comma, x expr = {\thisrow{Angle}},  y expr ={\thisrow{s_horz}}] {./tikz/GainPatterns30deg.csv};
	    \addplot[no markers, black, dashed] table [col sep=comma, x expr = {\thisrow{Angle}}, y expr ={\thisrow{GdB}}]{./tikz/gainPattern30deg.csv};
		
		\legend{\tiny{Gain}, \tiny{Vertically Polarized Pattern}, \tiny{Horizontally Polarized Pattern}, \tiny{Simulated Gain Pattern}}

		%\end{semilogxaxis}
	%	\legend{\footnotesize{Port 1}, \footnotesize{Port 2}, \footnotesize{Port 3}, \footnotesize{Port 4}}
		\end{polaraxis}
        %\draw (current bounding box.north east) -- (current bounding box.north west) -- (current bounding box.south) -- (current bounding box.center) -- cycle;
      %   \tikzstyle{every path}=[draw] 
         \path
            (current bounding box.center)coordinate(C)
            (current bounding box.east)coordinate(E)
            (current bounding box.north)coordinate(N)
            ;
         \pgfresetboundingbox
        \path let
              \p1=(C),
              \p2=(E),
              \p3=(N),
              \n1={\x2-\x1}, % diff center-east
              \n2={\y3-\y1}, % diff center-north
              \n3={\x1-\n1}, % New West
              \n4={\y1+1.1*\n2}, % New North
              \n5={\x1+\n1}, % New East
              \n6={0-1/6*\y3} % New south
              in 
              [use as bounding box] (\n3,\n4) rectangle (\n5,\n6);
            
	%	current bounding box.south = current bounding box.south+(5,0);
	%	\draw (current bounding box.north east) -- (current bounding box.north west) -- (current bounding box.south east) -- (current bounding box.south west) -- cycle;
\end{tikzpicture}
  \caption{30 degree}
\end{subfigure}
\begin{subfigure}[h]{0.65\textwidth}
  %{0.65\columnwidth}
\begin{tikzpicture}
    % \useasboundingbox (-1,-1.8) rectangle (15,8);
     %\draw [draw=black] (-1,-1.8) rectangle (15,8);
		\begin{polaraxis}[
		xmin = -90,
		xmax = 90,
		scale = 1,
		width = \columnwidth,
		xticklabel=$\pgfmathprintnumber{\tick}^\circ$,
		xtick={-90,-60,...,90},
	%	x coord trafo/.code=\pgfmathparse{#1+180},
	%	x coord inv trafo/.code= \pgfmathparse{#1-180},
	%	xtick={-90,-60,...,90},
		ytick={-10,-8,...,0},
		ymin=-10, ymax=1,
		y coord trafo/.code=\pgfmathparse{#1+10},
		rotate=-90,
		y coord inv trafo/.code=\pgfmathparse{#1-10},
		x dir=reverse,
		xticklabel style={anchor=-\tick-90},
		yticklabel style={anchor=east, xshift=2.5mm, yshift = -3mm},
		legend style={at={(0.6,0)}},
		legend columns = 4,
		ylabel = dB,
		y label style={at={(axis description cs: 0.54,0.18)},rotate=0},%,anchor=south},
	%	y axis line style={yshift=-4.75cm},
	%	ytick style={yshift=-4.75cm}
		]

		%,minGain,maxGain,meanGain,mpvarGain,mmvarGain
		\addplot[no markers, black] table [col sep =comma, x expr = {\thisrow{Angle}},  y expr ={\thisrow{G}}] {./tikz/GainPatterns45deg.csv};
		\addplot[no markers, blue] table [col sep =comma, x expr = {\thisrow{Angle}},  y expr ={\thisrow{s_vert}}] {./tikz/GainPatterns45deg.csv};
	    \addplot[no markers, green] table [col sep =comma, x expr = {\thisrow{Angle}},  y expr ={\thisrow{s_horz}}] {./tikz/GainPatterns45deg.csv};
	    \addplot[no markers, black, dashed] table [col sep=comma, x expr = {\thisrow{Angle}}, y expr ={\thisrow{GdB}}]{./tikz/gainPattern45deg.csv};
		
		\legend{\tiny{Gain}, \tiny{Vertically Polarized Pattern}, \tiny{Horizontally Polarized Pattern}, \tiny{Simulated Gain Pattern}}

		%\end{semilogxaxis}
	%	\legend{\footnotesize{Port 1}, \footnotesize{Port 2}, \footnotesize{Port 3}, \footnotesize{Port 4}}
		\end{polaraxis}
		 %\draw (current bounding box.north east) -- (current bounding box.north west) -- (current bounding box.south) -- (current bounding box.center) -- cycle;
       %  \tikzstyle{every path}=[draw] 
         \path
            (current bounding box.center)coordinate(C)
            (current bounding box.east)coordinate(E)
            (current bounding box.north)coordinate(N)
            ;
         \pgfresetboundingbox
        \path let
              \p1=(C),
              \p2=(E),
              \p3=(N),
              \n1={\x2-\x1}, % diff center-east
              \n2={\y3-\y1}, % diff center-north
              \n3={\x1-\n1}, % New West
              \n4={\y1+1.1*\n2}, % New North
              \n5={\x1+\n1}, % New East
              \n6={0-1/6*\y3} % New south
              in 
              [use as bounding box] (\n3,\n4) rectangle (\n5,\n6);
            
	%	current bounding box.south = current bounding box.south+(5,0);
	%	\draw (current bounding box.north east) -- (current bounding box.north west) -- (current bounding box.south east) -- (current bounding box.south west) -- cycle;
		
\end{tikzpicture}
  \caption{45 degree}
\end{subfigure}
\caption{Results of simulation and measurement employing different precoding vectors calculated for omnidirectional radiation and a maximal gain at 30 and 45 degree.}
\label{fig:MeasurementResults}
\end{figure}
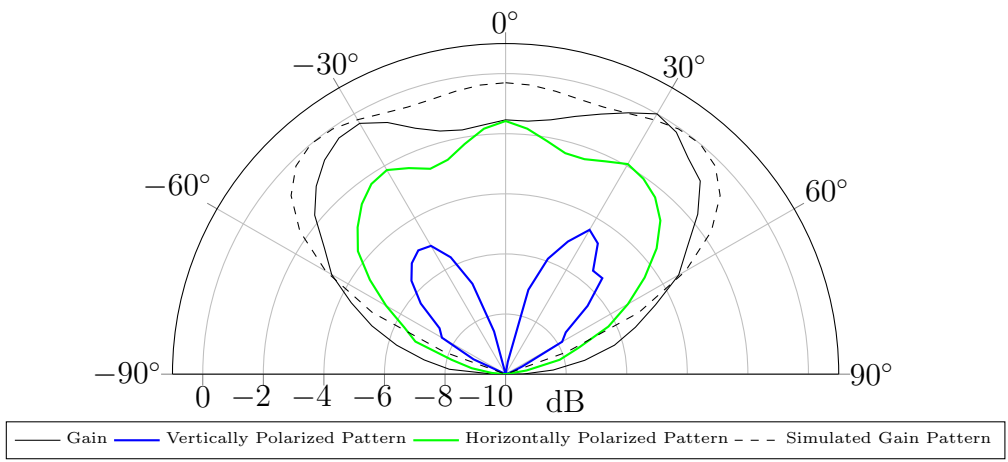
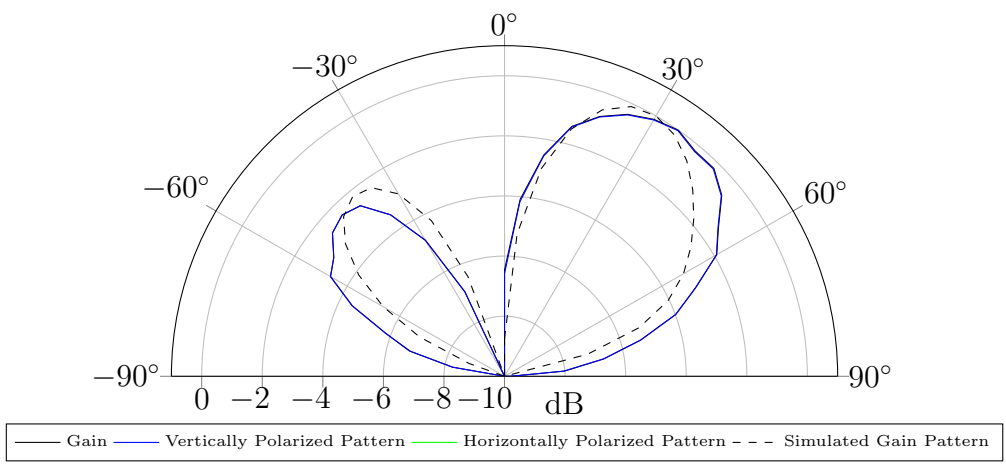
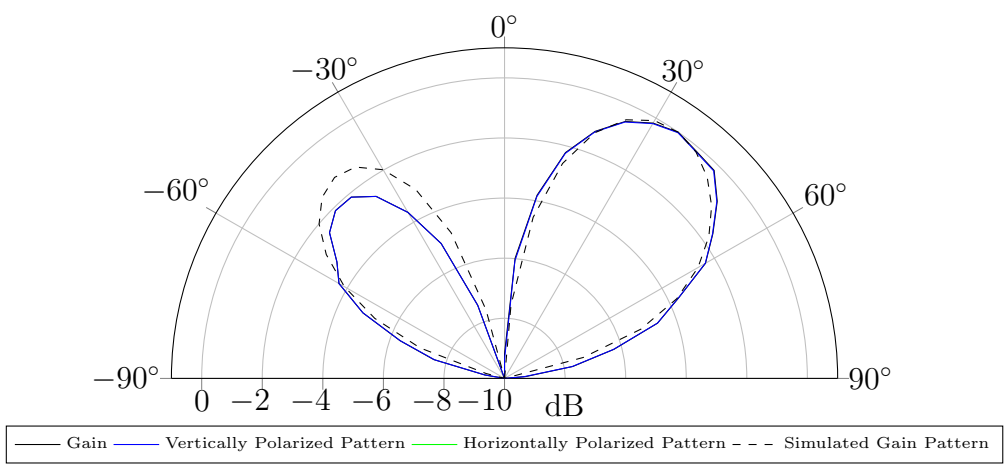

In the context of the measurements, three different sets of coefficients have been tested.
The first one is optimized to achieve a close to omnidirectional radiation in the angular region between $-45^{\circ}$ to $45^{\circ}$, with $0^{\circ}$ being the broadside direction.
The two other measurements deal with the optimization of the gain towards a specific angle, namely
$30^{\circ}$ and $45^{\circ}$.
The numbers of the ports used at the antenna are number 2, 4, 5, and 6.
The codebook entries are given in table \ref{tab:CodebookEntries}.
\begin{table}[h]
    \centering
    \caption{Codebook entries for the measurements.}
    \label{tab:CodebookEntries}
    \begin{tabular}{c|c|c|c}
    Port Number & Omnidirectional & 30 degree & 45 degree \\
    \hline
    2 &  $0.23$ & $0.5250$ & $0.7555$ \\
    4 &  $-0.60-j0.02$ & $j0.7607$ & $j0.6070$ \\
    5 &  $0.77$ & $0.0001+j0.0001$ & $0.0003+j0.0001$ \\
    6 &  $0$ & $0.3818$ & $0.2465$ \\
    \end{tabular}
\end{table}
The optimization results are quite similar for both optimizations for 30 and 45 degrees, although the precoding vectors are different, which can be seen from the dashed lines in Fig.~\ref{fig:MeasurementResults}.
As can be seen in Fig.~\ref{fig:MeasurementResults}, the measurements show good fit with the previously calculated radiation patterns of the theoretical antenna model.
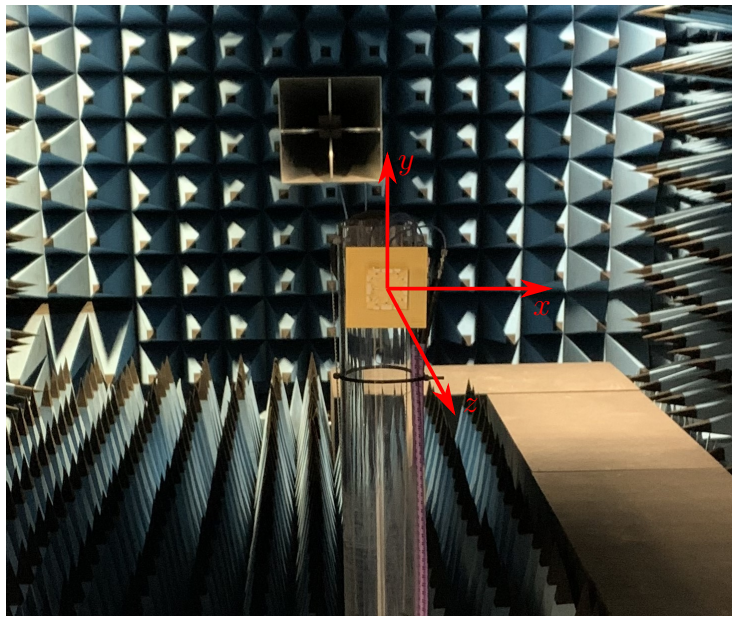
\begin{figure}[h]
    \centering
    \def\svgwidth{0.6\textwidth}
    %% Creator: Inkscape 1.1.2 (b8e25be833, 2022-02-05), www.inkscape.org
%% Accompanies image file 'AntCut.pdf' (pdf, eps, ps)
\begingroup%
  \makeatletter%
  \providecommand\color[2][]{%
    \errmessage{(Inkscape) Color is used for the text in Inkscape, but the package 'color.sty' is not loaded}%
    \renewcommand\color[2][]{}%
  }%
  \providecommand\transparent[1]{%
    \errmessage{(Inkscape) Transparency is used (non-zero) for the text in Inkscape, but the package 'transparent.sty' is not loaded}%
    \renewcommand\transparent[1]{}%
  }%
  \providecommand\rotatebox[2]{#2}%
  \newcommand*\fsize{\dimexpr\f@size pt\relax}%
  \newcommand*\lineheight[1]{\fontsize{\fsize}{#1\fsize}\selectfont}%
  \ifx\svgwidth\undefined%
    \setlength{\unitlength}{1090.5bp}%
    \ifx\svgscale\undefined%
      \relax%
    \else%
      \setlength{\unitlength}{\unitlength * \real{\svgscale}}%
    \fi%
  \else%
    \setlength{\unitlength}{\svgwidth}%
  \fi%
  \global\let\svgwidth\undefined%
  \global\let\svgscale\undefined%
  \makeatother%
  \begin{picture}(1,0.84731774)%
    \lineheight{1}%
    \setlength\tabcolsep{0pt}%
    \put(0,0){\includegraphics[width=\unitlength,page=1]{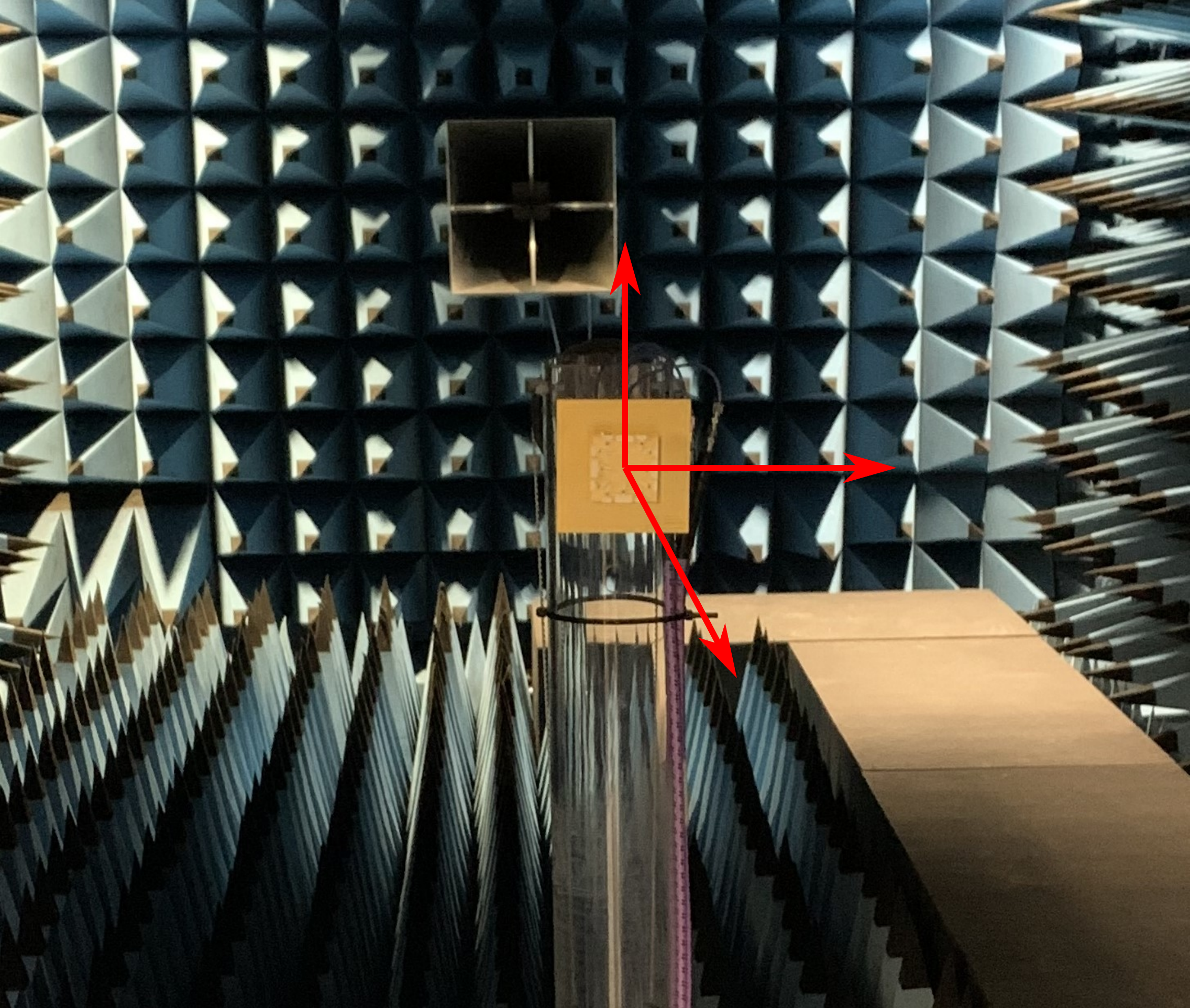}}%
    \put(0.72632684,0.41757118){\color[rgb]{0,0,0}\makebox(0,0)[lt]{\lineheight{1.25}\smash{\begin{tabular}[t]{l}\textcolor{red}{${x}$}\end{tabular}}}}%
    \put(0.62915858,0.28219835){\color[rgb]{0,0,0}\makebox(0,0)[lt]{\lineheight{1.25}\smash{\begin{tabular}[t]{l}\textcolor{red}{${z}$}\end{tabular}}}}%
    \put(0.54034928,0.61986709){\color[rgb]{0,0,0}\makebox(0,0)[lt]{\lineheight{1.25}\smash{\begin{tabular}[t]{l}\textcolor{red}{${y}$}\end{tabular}}}}%
  \end{picture}%
\endgroup%

    \caption{Coordinate system of the antenna during the measurement. The pattern is measured in the $x$-$z$-plane.}
    \label{fig:my_label}
\end{figure}
\newpage
\vfill
\vfill
\clearpage

\section{Conclusion}
The results have shown good fit between the optimized radiation patterns of the model of the antenna and the real antenna.
Hence, optimization and application of beamforming using the knowledge of the radiation of a multi-mode antenna element is possible.

\pagebreak
\bibliography{IEEEabrv.bib,ICTabrv.bib,literature.bib}
\bibliographystyle{IEEEtran}
\pagebreak
\section*{Appendix}
In this section, conducted measurement results are provided which have not been used for the setup of single-antenna-element beamforming.

\subsection*{Measurement of Transmission Lines}
As preparation for alternative calibration procedures, the phase differences and attenuations at a frequency $f=2~\textrm{GHz}$ of some SMA-cables have been measured, using a VNA R\&S ZVA 40. The results are listed in table~\ref{tab:SMA_Cable}.
\begin{table}[h]
    \centering
    \caption{Phases of transmission lines at frequency $f=2~\textrm{GHz}$}
    \label{tab:SMA_Cable}
    \begin{tabular}{c|c|c}
    
        Transmission Line Serial Number & Attenuation in $dB$ & Phase in degree \\
         \hline
        189204 & 0.71 & 128.7\\
        189230 & 0.65 & 136.9\\
        189236 & 0.71 & 103.4\\
        189237 & 0.69 & 95.4
    \end{tabular}

\end{table}
%\printbibliography % Be sure to remove access date and months from the .bib file

\begin{figure}[t]
    \centering
    \includegraphics[page=1,width=\textwidth]{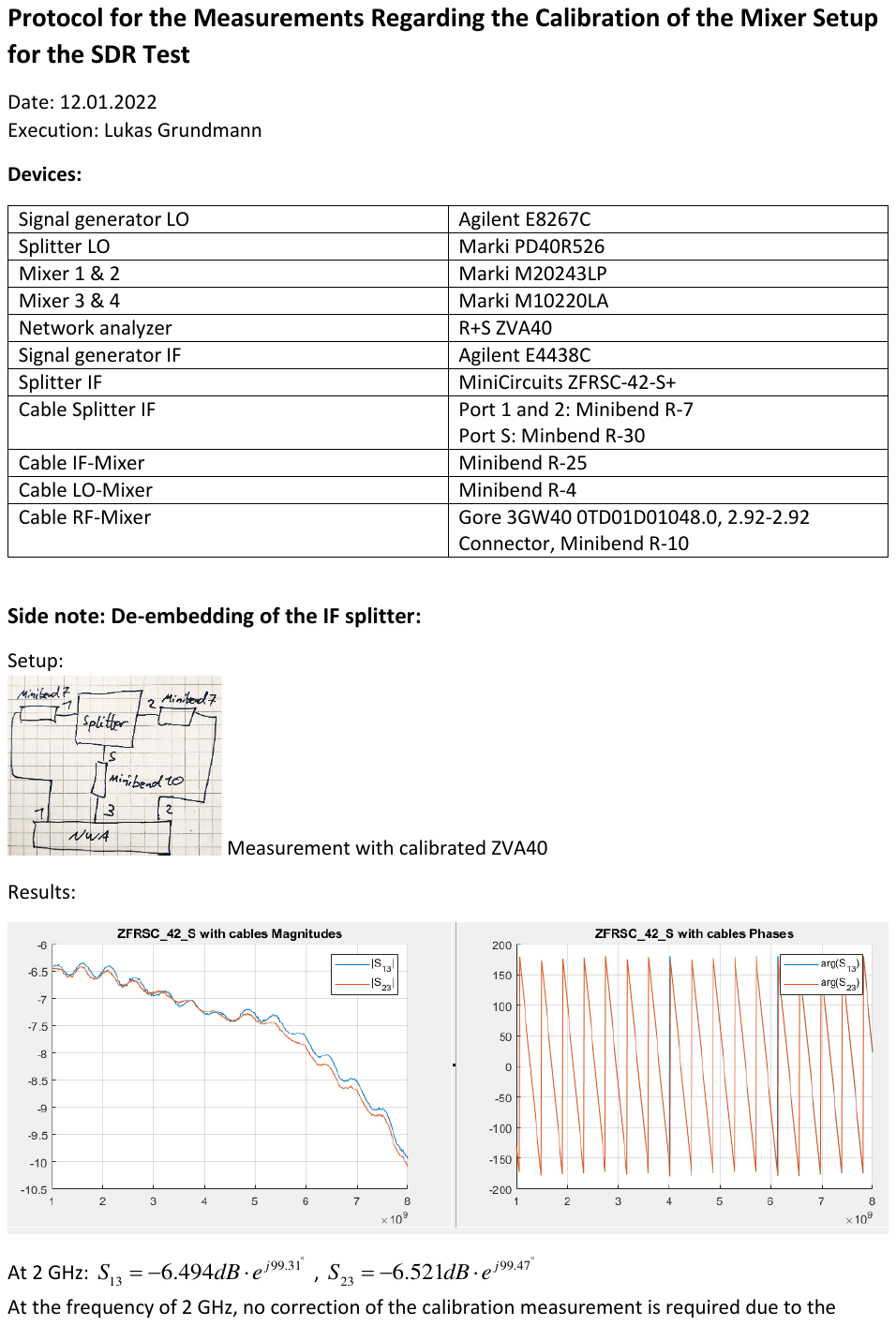}
\end{figure}
\begin{figure}[t]
    \centering
    \includegraphics[page=2,width=\textwidth]{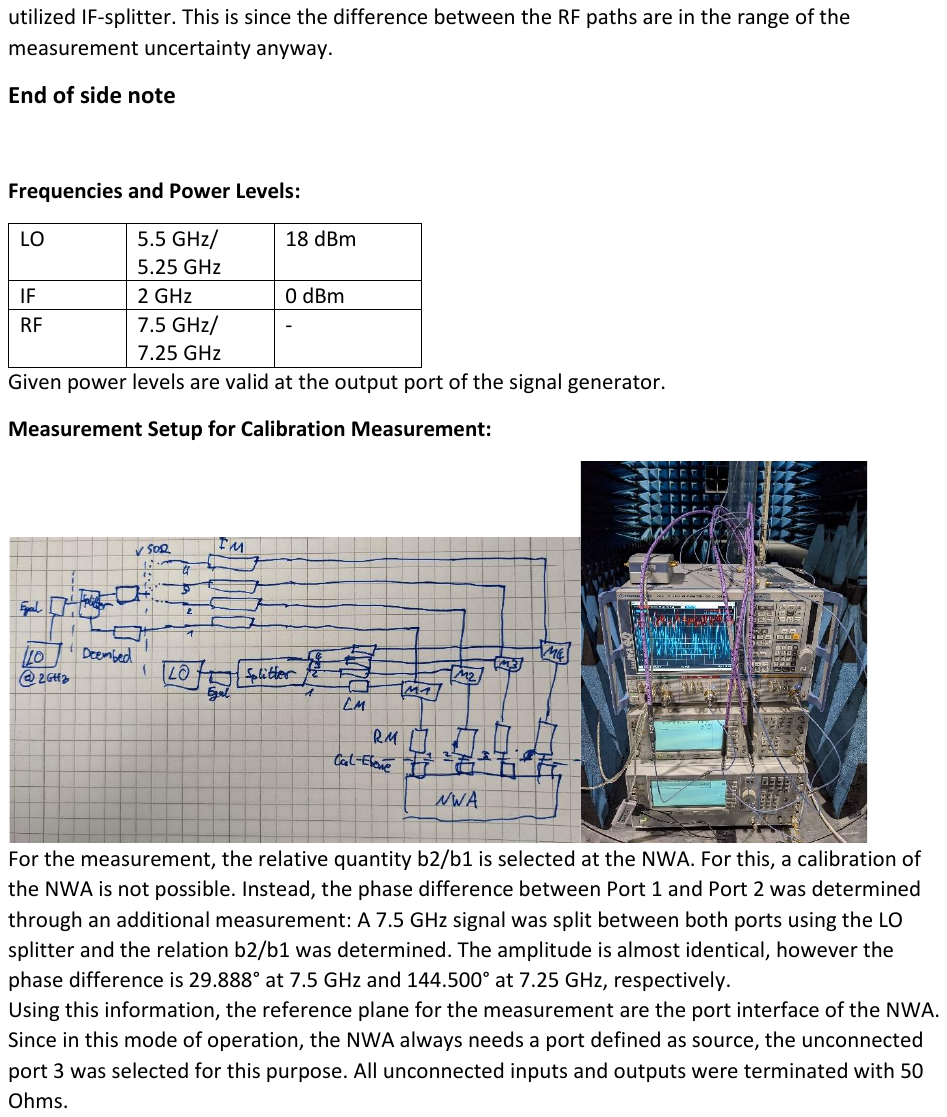}
\end{figure}
\begin{figure}[t]
    \centering
    \includegraphics[page=3,width=\textwidth]{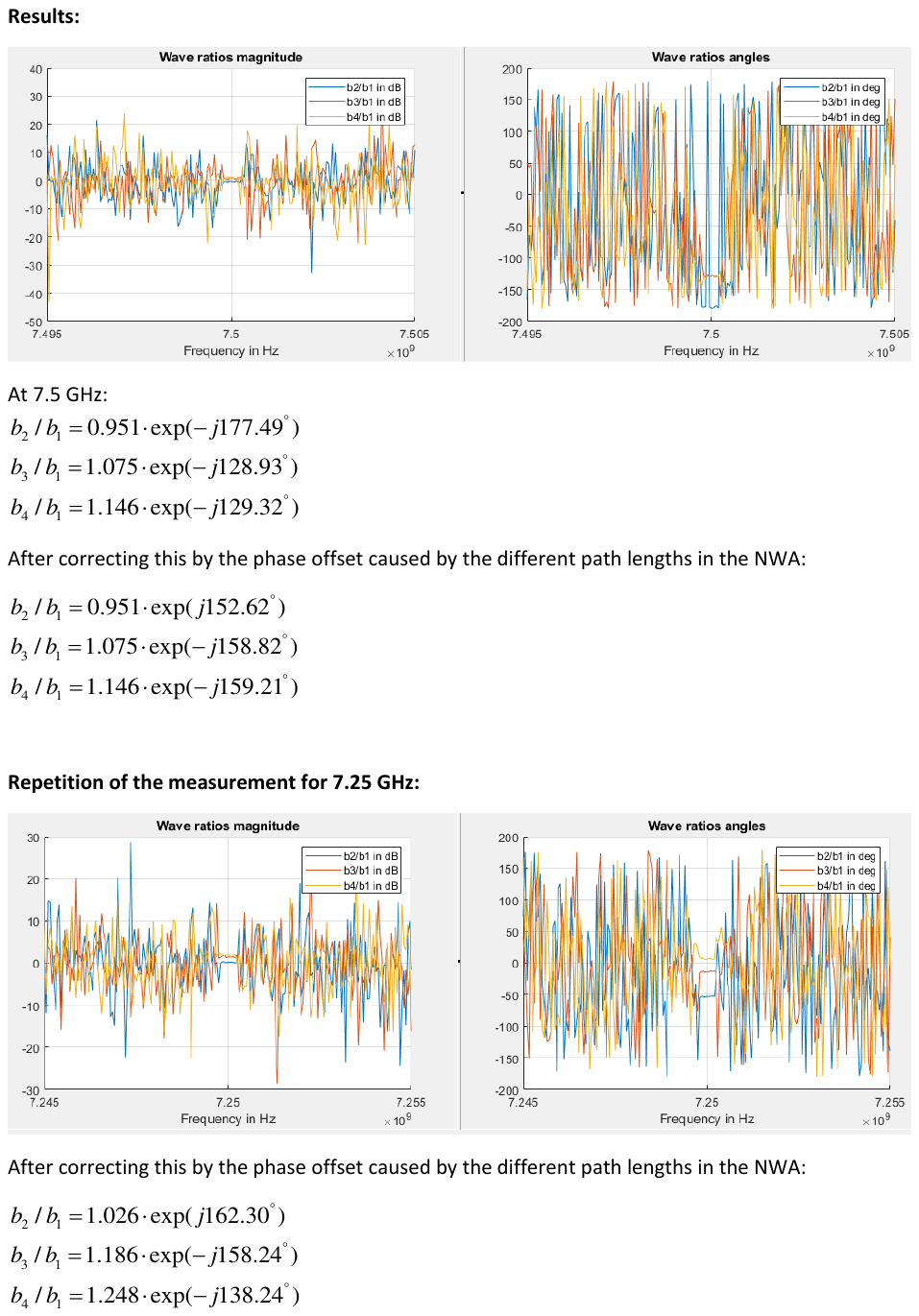}
\end{figure}
\begin{figure}[t]
    \centering
    \includegraphics[page=4,width=\textwidth]{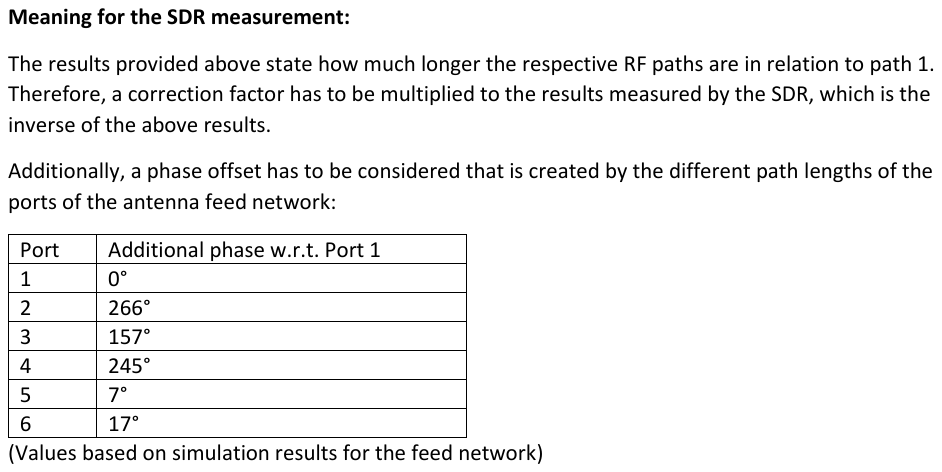}
\end{figure}

%\includepdf[pages={1-4},scale=1]{./tikz/Protokoll_Cal_Measurement_en}

\end{document}